\documentclass[12pt]{article}
\usepackage{graphics}
\begin{document}
\begin{center}
{The Non-perturbative Interaction of the Pseudovector Coupling \\ for the Pion-nucleon Scattering}
\end{center}
\begin{center}
{Susumu Kinpara}
\end{center}
\begin{center}
{\it National Institute of Radiological Sciences \\ Chiba 263-8555, Japan}
\end{center}
\begin{abstract}
The effect of the non-perturbative term is investigated for the pion-nucleon scattering. 
Including the self-energy of nucleon 
the cross section of the elastic scattering results in a constant value at the laboratory momentum $p_L \to \infty$.
The amplitude of the forward direction is applied to the dispersion relation.
\end{abstract}
\section*{\normalsize{1 \quad Introduction}}
\hspace*{4.mm}
The dynamics of the constituents such as nucleons and mesons are the basis of many nuclear phenomena.
The lightest meson is the pion having the longest range of the interaction and which is concerned with the nuclear system
in the bound and the scattering states primarily.
It is our interest to elucidate the properties in the framework of the quantum field with the symmetries
of the composite system.
\\\hspace*{4.mm}
To investigate the meson-exchange model the interaction of pion with nucleon is indispensable 
and at present it is divided into two types that is the pseudoscalar and the pseudovector interactions.
Since empirically the latter is favorable for the calculation of the finite nuclei 
and the nucleon-nucleon elastic scattering by the Bethe-Salpeter equation we use it to proceed the present study.
It will be pointed out later that both interactions are related with each other by the non-perturbative term
arising from the pseudovector interaction under the breaking of the chiral symmetry. 
\\\hspace*{4.mm}
Recent our study has revealed 
that the nucleon propagator plays a decisive role for the electromagnetic form factor \cite{Kinpara}.
The explicit form of the self-energy is on the basis of the perturbative expansion 
together with the counter terms for the mass and the wave function. 
It is interesting that the lowest-order form 
is independent of the pseudovector coupling constant and consequently makes us possible to adjust the strength of the effect.  
\\\hspace*{4.mm}
The lowest-order perturbative calculation does not work well 
for the electromagnetic properties of proton because the effect is too strong to use it 
unless the terms of the higher-order in the self-energy is approximately dropped.
Therefore the present formulation needs the diagrams of the higher-order processes to interpret the system correctly.
\\\hspace*{4.mm}
The pion-nucleon scattering is an important process
to examine whether the results taking account of the self-energy follow the trends
in the electromagnetic properties or not.
Without the non-perturbative term the calculation of the cross section shows the divergent behavior 
as the incident energy of pion grows and it breaks the unitary bounds obviously.
To recover from the lack of the suppression the pseudovector interaction should be supplemented by the additional force.
Our purpose of the present study is to examine the effect on the pion-nucleon scattering.
\\
\section*{\normalsize{2 \quad The non-perturbative term for the elastic scattering }}
\hspace*{4.mm}
The usual vertex of the pseudovector coupling interaction is corrected 
by the non-perturbative term 
to construct the pion-nucleon-nucleon three-point vertex.
In our previous study the extended one has been applied to the electromagnetic properties of nucleon
such as the anomalous magnetic moment and the electromagnetic form factor of proton \cite{Kinpara}. 
The present case of the pion-proton scattering is similar to them and the vertex $\Gamma_0(p,q)$ is modified as
\begin{eqnarray}
\Gamma_0(p,q)
\rightarrow
\Gamma_0(p,q)+\Gamma_1(p,q)
\end{eqnarray}
\begin{eqnarray}
\Gamma_1(p,q) = G(p)^{-1} \, \gamma_5 \, +  \, \gamma_5 \, G(q)^{-1} 
\end{eqnarray}
between the incoming ($\it q$) and the outgoing ($\it p$) four-momenta of proton.
The isospin matrix is omitted for the sake of brevity.
The $\Gamma_0(p,q)$ is the vertex of the pseudovector coupling interaction.
When the lowest-order approximation is used it is given as $\Gamma_0(p,q) = \gamma_5 \gamma\cdot (p-q)$.
In Eq. (2) the inverse of the nucleon propagator $G(p)^{-1}= \gamma\cdot p - M - \Sigma (p)$
contains the self-energy $\it\Sigma(p)$ also in free space.
An interesting character is that when $\Sigma(p) \equiv 0$ the extended interaction 
is identical with that of the pseudoscalar coupling.
Since $\it\Sigma(p)$ is expanded in powers of $\gamma\cdot p - M$ the difference 
between two types of the interactions
appears in the intermediate state for lack of the on-shell property of the four-momentum.
\\\hspace*{4.mm}
The general form of $\it\Sigma(p)$ is expressed by $\Sigma(p) = c_1(p^2) - \gamma\cdot p \, c_2(p^2)$ 
in terms of the coefficients $c_1(p^2)$ and $c_2(p^2)$.
These are determined taking into account the non-perturbative relation together with the counter terms.
The exact form of $\it\Sigma(p)$ is defined as the summation of the series in $\gamma\cdot p - M$ up to the infinite order.
While the effect of the self-energy term is too strong to describe the anomalous magnetic moment 
the lowest-order approximation works well.
In fact the experimental value is reproduced by
the calculation of the perturbative expansion for both of the pion and the nucleon currents.
The higher-order corrections in the self-energy may change the structure of the power series 
so as to suppress the higher-order terms.
As will be seen later the exact form is essential to derive the cross section of the elastic scattering
at the high energy limit.
\\\hspace*{4.mm}
Applying the non-perturbative term to the pion-proton scattering the part of the fermion line in the scattering amplitude
for $\pi^{+}$-proton case is extended to include it as
\begin{eqnarray}
(\Gamma_0(p^\prime,p-k^\prime)+\Gamma_1(p^\prime,p-k^\prime))\, G(p-k^\prime)
\,(\Gamma_0(p-k^\prime,p)+\Gamma_1(p-k^\prime,p)) \nonumber\\ 
= \alpha_1 + \alpha_2 + \alpha_3 + \alpha_4
\end{eqnarray}
\begin{eqnarray}
\alpha_1 &\equiv& \Gamma_0(p^\prime,p-k^\prime) \, G(p-k^\prime) \, \Gamma_0(p-k^\prime,p)\nonumber\\
&\approx&
\Gamma_0(p^\prime,p-k^\prime) \, G_0(p-k^\prime) \, \Gamma_0(p-k^\prime,p)
\end{eqnarray}
\begin{eqnarray}
\alpha_2 \;\equiv\; \Gamma_0(p^\prime,p-k^\prime) \, G(p-k^\prime) \, \Gamma_1(p-k^\prime,p) = -\gamma \cdot k
\end{eqnarray}
\begin{eqnarray}
\alpha_3 \;\equiv\; \Gamma_1(p^\prime,p-k^\prime) \, G(p-k^\prime) \, \Gamma_0(p-k^\prime,p) = -\gamma \cdot k^\prime
\end{eqnarray}
\begin{eqnarray}
\alpha_4 &\equiv& \Gamma_1(p^\prime,p-k^\prime) \, G(p-k^\prime) \, \Gamma_1(p-k^\prime,p) \nonumber\\
&=& \gamma \cdot k^\prime -2 M - \gamma_5 \Sigma(p-k^\prime) \gamma_5
\end{eqnarray}
\\
in which $p$, $k$, $p^\prime$ and $k^\prime$ are the initial four-momenta of proton ($p$) and pion ($k$)
and the final four-momenta of proton ($p^\prime$) and pion ($k^\prime$).
The momenta are connected by the relation of the conservation as $p + k = p^\prime + k^\prime$.
Deriving Eqs. (4)$\sim$(7) these quantities $\alpha_i$ ($i\,$=$\,$1,$\cdots$,4) 
are assumed to be sandwitched by the Dirac spinors
from both sides as $\bar{u}^{(s^\prime)}(p^\prime)\,\sum_{i=1,\cdots ,4}\alpha_i\,u^{(s)}(p)$.
\\\hspace*{4.mm}
When there is not the non-perturbative term only the part $\alpha_1$ remains and it constructs the scattering amplitude 
within the Born approximation by the pseudovector coupling interaction.
In Eq. (4) the exact nucleon propagator is approximated 
to the free one $G_0(p-k^\prime)=(\gamma\cdot (p-k^\prime) -M +i \epsilon)^{-1}$ without the term of the self-energy.
The cross section $\sigma$ of the elastic scattering by the lowest-order perturbative expansion 
results in $\sigma \rightarrow \infty$ rapidly as the incident energy increases and it contradicts the experimental fact.
\\\hspace*{4.mm} 
The part $\alpha_2+\alpha_3+\alpha_4$ is calculated to correct the part $\alpha_1$ 
instead of the inclusion of the self-energy in $\alpha_1$ and the higher-order diagrams.
Using Eqs. (3)$\sim$(7) the amplitude of the $\pi^{+}$-proton scattering is    
\begin{eqnarray}
M^{(s,s^\prime)}(p,p^\prime,k,k^\prime) \,=\, \bar{u}^{(s^\prime)}(p^\prime)\,\hat{T}\,u^{(s)}(p)
\end{eqnarray}
\begin{eqnarray}
\hat{T} \,=\, A(x)-\frac{\gamma\cdot (k+k^\prime)}{2}\,B(x)
\end{eqnarray}
\begin{eqnarray}
A(x)\,\equiv\, -\frac{g^2}{M}\,C(x)
\end{eqnarray}
\begin{eqnarray}
B(x)\,\equiv\, 2 g^2 x^{-1} \,(1-\frac{x}{2 M} D(x))
\end{eqnarray}
in terms of $C(x)$ and $D(x)$ which represent the effect of the self-energy in $\alpha_4$.
The argument $x \equiv -(p-k^\prime)^2 + M^2 = 2 p \cdot k^\prime - m_\pi^2$ 
is the denominator of the rationalized propagator of fermion.
The coupling parameter $g$ is defined as $g \equiv 2 M f_\pi/m_\pi$ where $f_\pi$ and $m_\pi$ are
the pseudovector coupling constant and the mass of pion.
When $C(x) = D(x) \equiv 0 $ the amplitude $M^{(s,s^\prime)}(p,p^\prime,k,k^\prime)$ is equivalent to the result
of the pseudoscalar coupling interaction.
In the case of the $\pi^{-}$-proton scattering the amplitude is obtained by the replacement $k \leftrightarrow - k^\prime$
by virtue of the crossing symmetry.
\\\hspace*{4.mm}
The self-energy of nucleon has been calculated by the lowest-order perturbative treatment 
and then it does not depend on the coupling constant $f_\pi$.
The singular property is favorable to adjust the magnetic moment to the experimental value. 
The explicit forms of the functions $C(x)$ and $D(x)$ are 
\begin{eqnarray}
C(x) = \frac{2 M^2 (M^2+m_\pi^2)-m_\pi^2 x/2-x^2/4}{m_\pi^2 (M^2+m_\pi^2)+(3M^2/4+m_\pi^2)x+x^2/4}
\end{eqnarray}
\begin{eqnarray}
D(x) = \frac{M(M^2+m_\pi^2)+M x/4}{m_\pi^2 (M^2+m_\pi^2)+(3M^2/4+m_\pi^2)x+x^2/4}.
\end{eqnarray}
At the high energy limit $x \rightarrow \infty$, they approach to the values $C(x)\rightarrow -1$ and $D(x)\rightarrow 0$
respectively. 
It resembles to the constant values $C(x) = -1$ and $D(x) = (2M)^{-1}$
of the pseudovector interaction without the non-perturbative term when the target is infinitely heavy ($M \to \infty$). 
\\\hspace*{4.mm}
Besides the pole term resulting from the pseudoscalar interaction, 
the other two poles in the $C(x)$ and $D(x)$ are not realistic and it is difficult to understand the numerical result.
Particularly the two poles are at $x < 0$ and one of which is in the physical region of the pion energy.
The experimental data of the $\pi^{-}$-proton scattering encountering no pole structure there
indicate that the application of the exact form of the self-energy is restricted to the high energy region.
\\\hspace*{4.mm}
As seen from the study of the electromagnetic form factor the approximate form of the self-energy
is useful to investigate the properties of the scattering around the intermediate energy region.
In the lowest-order approximation the self-energy becomes $\Sigma(p) \approx M\,(M^2+m_\pi^2)^{-1}(\gamma\cdot p -M)^2$ 
and the coefficients are given by $C(x) \approx (2M^2-x/2)(M^2+m_\pi^2)^{-1}$ and $D(x) \approx M (M^2+m_\pi^2)^{-1}$
having no poles.
The cross section is divergent as the energy increases like the case of the pseudovector coupling
without the non-perturbative term.
\\\hspace*{4.mm}
Choosing the set of $C(x)$ and $D(x)$ the cross section of the elastic scattering $\sigma$ is obtained 
by integrating the differential cross section
\begin{eqnarray}
\frac{d \sigma}{d \Omega}=\frac{M^2}{32 \pi^2 s}\,
\sum_{s, \, s^\prime = \pm 1/2}\vert M^{(s,s^\prime)}(p,p^\prime,k,k^\prime) \vert^2
\end{eqnarray}
over the solid angle in the center of mass system.
The $s$ in front of the summation is the Lorentz invariant quantity $s \equiv (p+k)^2$.
\\\hspace*{4.mm}
Because the self-energy is intact under the change of $f_\pi$ within the present approximation
the high energy limit of the cross section of the elastic scattering denoted by $\sigma_\infty$ 
is in the simple relation with $f_\pi^2$.
For example in the case of $\pi^{-}$-proton elastic scattering it is given by
\begin{eqnarray}
f_\pi^2 = \frac{m_\pi^2}{M}\sqrt{2 \pi \sigma_\infty}.
\end{eqnarray}
The experimental value is about $\sigma_\infty \sim $ 3.3 mb at the laboratory beam momentum 100 GeV/c.
Therefore by using Eq. (15) it yields $f_\pi \sim 0.4$ which is smaller than the standard value $f_\pi \sim 1$.
\\
\section*{\normalsize{3 \quad The non-perturbative term for the total cross section }}
\hspace*{4.mm}
To understand the phenomena at the intermediate energy ($\leq$ 1 GeV) is our main interest.
The perturbative method is not efficient to explain the process particularly for the low energy region.
For example below the resonance energy 
the $S$-wave is dominant and the repulsive force is strong so that the result of the calculation overestimates the experimental values.
We need to use the non-perturbative method such as the dispersion relation for the pion-nucleon scattering
in conjunction with the unitarity of the S-matrix to obtain the total cross section \cite{BD}.
\\\hspace*{4.mm}
The dispersion relation connects the real and the imaginary parts of the forward scattering amplitude $T(\omega)$ 
as follows:
\begin{eqnarray}
T(\omega) \equiv \frac{1}{2} \sum_{s=\pm 1/2} M^{(s,s)}(p,p,k,k) = A(\omega) +\omega B(\omega)
\end{eqnarray}
\begin{eqnarray}
{\rm Re} \, T(\omega) = \sum_i \frac{a_i}{\omega - \omega_i} 
+\frac{{\rm P}}{\pi}\{\int_{-\infty}^\infty - \int_{-m_\pi}^{m_\pi} \} \,
d \omega^\prime \, \frac{{\rm Im}\,T(\omega^\prime)}{\omega^\prime-\omega}
\end{eqnarray}
in terms of the pion laboratory energy $\omega \equiv k\cdot p/M$.
The P denotes the principal value of the integral.
When there is not the self-energy term or at least it does not contain pole terms within the contour circle
the summation in Eq. (17) consists of the single term with the location $\omega_i \equiv -m_\pi^2/2M$ 
and the residue $a_i \equiv g^2 m_\pi^2/2M^2$
arising from the nucleon propagator.
\\\hspace*{4.mm}
The pole terms possibly appear due to the self-energy on the real axis 
other than the region between $-m_\pi$ and $m_\pi$. 
In such a case the delta functions are added to the solution for $Im \, T(\omega)$.
The exact forms $C(\omega)$ and $D(\omega)$ in Eq. (12) and (13) actually have such a pole outside the circle.
Moreover application of it does not give results explaining properly the energy dependence of the total cross section.
Then we proceed our calculations by using the approximate lowest-order form encountering no external poles.
\\\hspace*{4.mm}
Eq. (17) is used to examine the role of the non-perturbative term.
The second term in the curly brackets is approximated by the simple form
\begin{eqnarray}
- \frac{1}{\pi}\int_{-m_\pi}^{m_\pi} \,
d \omega^\prime \, \frac{{\rm Im}\,T(\omega^\prime)}{\omega^\prime-\omega}
\approx a\,\omega^{-1} \;\; (\;a \equiv \pi^{-1}\int_{-m_\pi}^{m_\pi} d \omega \, {\rm Im}\,T(\omega)\;)
\end{eqnarray}
at $\vert\omega\vert > \alpha > m_\pi$.
It becomes the delta function $\sim \delta(\omega)$ same as the external pole.
The parameter $\alpha$ is introduced to limit the region in which the model of $Re \,T(\omega)$ is applied to carry out
the following integral over $\omega^\prime$.
\\\hspace*{4.mm}
Removing the pole terms and assuming the Hilbert transform the $Im\,T(\omega)$ is expressed by the integral form
\begin{eqnarray}
{\rm Im} \, T(\omega) = -\frac{{\rm P}}{\pi} \int_{-\infty}^\infty \,
d \omega^\prime \, \frac{{\rm Re}\,T(\omega^\prime)}{\omega^\prime-\omega}.
\end{eqnarray}
Consequently the real part of $T(\omega)$ is related to the total cross section $\sigma_{\pm}(\omega)$ 
of the $\pi^{\pm}$-proton scattering according to the relation about the unitarity of the S-matrix
\begin{eqnarray}
\sigma_{\pm}(\omega) = \frac{\mp 1}{\sqrt{\omega^2-m_\pi^2}}\,{\rm Im} \, T(\mp \omega).
\end{eqnarray}
\\\hspace*{4.mm}
Our next task is to prepare $Re \, T(\omega)$ appropriate to perform the transform.
Up to the linear term within the lowest-order perturbative calculation it is expressed by
\begin{eqnarray}
{\rm Re} \, T(\omega) = -\frac{g^2}{M}(\, b_0 + b_1\,\frac{\omega}{M} \,)
\end{eqnarray}
where the pole term is left out since it contributes to the part of the delta function.  
The coefficients $b_0$ and $b_1$ in four models of the pseudovector interaction are tabulated in Table 1 
neglecting the terms of the order $O(m_\pi^2/M^2)$ much smaller than the zeroth order.
\begin{center}
\begin{tabular}{|c|c c c c|}
   \multicolumn{5}{c}{ Table 1 }\\
      \hline
 & no $\Gamma_1$ & no $\Sigma$ & lowest $\Sigma$ & $\Sigma_\infty$ \\
     \hline
$\quad b_0 \quad$ & 0 & 1 & 3 & $-$1/2 \\
     \hline
$\quad b_1 \quad$ & 1/2 & 0 & 2 & 0 \\
     \hline
\end{tabular}
\end{center}
The column $no \,\it\Gamma_1$ is the model for which the non-perturbative term $\Gamma_1(p,q)$ in Eq. (2) is turned off
that is the usual vertex of the pseudovector coupling interaction.
The other three models include the non-perturbative term.
The column $no \,\it\Sigma$ is the model the self-energy $\it\Sigma(p)$ is set equal to zero 
and then the vertex is identical with that of the pseudoscalar interaction.
The column $lowest \,\it\Sigma$ uses the approximate self-energy by the lowest-order calculation.
The coefficients using the exact form is in the column $\it\Sigma_\infty$.
\\\hspace*{4.mm}
When only the constant term ($\sim b_0$) exists the integral of Eq. (19) is carried out over $\omega^\prime$ 
at the region $(\,-\infty,\, \alpha\,]$ and $[\,\alpha,\, \infty\,)$ and it becomes
\begin{eqnarray}
{\rm Im} \, T(\omega) = \frac{g^2 b_0}{\pi M}\, \log \left| \frac{\omega+\alpha}{\omega-\alpha} \right|
\end{eqnarray}
along with the parameter $\alpha$ owing to the uncertainty of $Re\,T(\omega)$ at ($-\alpha$,$\,$$\alpha$).
The $Im \,T(\omega)$ in Eq. (22) is divergent at $\omega = \alpha$ 
and the value of $\alpha$ is determined so as to adjust the resonance energy of the $\pi$-proton scattering 
which is $\alpha\sim$ 0.3 GeV.
Since $Im \,T(\omega)$ is anti-symmetric ($\,{\rm Im} \,T(-\omega)$ = $-{\rm Im} \,T(\omega)\,$)
the constant term alone does not cause the difference of the cross section 
between $\sigma_{+}(\omega)$ and $\sigma_{-}(\omega)$.
The sign of $b_0$ is significant and required to be positive.
The model $\it\Sigma_\infty$ has the negative value
and therefore not suitable for the present approach of the dispersion method
although the perturbative calculation of the cross section for the elastic scattering 
gives the constant value at the high-energy limit.
\\\hspace*{4.mm}
The anti-symmetric component in $Re \,T(\omega)$ is expected to interpret the separation of $\sigma_\pm$
which is observed by the experiment as $\sigma_{+}/\sigma_{-}\sim 3$ around the resonance energy $\omega\sim 0.3$ GeV.
To incorporate the linear term being divergent at $\omega \to \pm \infty$ in the transform 
the variable $\omega$ is replaced with the form of the Fourier expansion as
\begin{eqnarray}
\omega \to \frac{2 L}{\pi} \sum_{n=1}^\infty \frac{(-1)^{n-1}}{n} \,{\rm sin} \frac{n \pi \omega}{L}.
\end{eqnarray}
It is a periodic function identical to $\omega$ at $-L<\omega<L$.
The Hilbert transform is applied to each term and the summation results in the additional part
\begin{eqnarray}
{\rm Im} \, T(\omega) = \frac{L g^2 b_1}{\pi M^2}\, \{\log 2 + \log (1+{\rm cos}\frac{\pi \omega}{L})\}
\end{eqnarray}
where the final result is dependent on the parameter $L$.
For the quantity $Im \, T(\omega)$ in Eq. (24) is divergent at the upper limit $\omega = L$
the value of $L$ is required to be greater than the intermediate energy region so that the total cross section 
is explained by the present model.
\\\hspace*{4.mm}
The coefficients $b_1$ of the four models in Table 1 all have the non-negative values 
and the respective value of $\sigma_{-}$ is found to be larger than that of $\sigma_{+}$ 
at the intermediate energy region contrary to the experimental fact.
It is reasonable to proceed calculation of the self-energy next to the lowest-order 
to correct the model $lowest \,\it\Sigma$.
For the present calculation the $\sim \omega^2$ term is neglected 
which is symmetric as well as the constant term and moves $\sigma_\pm$ toward the same direction.
The coefficients of the constant and the linear terms shift from those of the lowest-order
as $b_0 = 3 \to 1$ and $b_1 = 2 \to -1$ by the inclusion of the second-order calculation.
The sign of $b_1$ turns to negative in agreement with the sizes of $\sigma_\pm(\omega)$.
\\\hspace*{4.mm}
The set of $(b_0,b_1) = (1,-1)$ is the second-order approximation to the exact one.
In comparison with the other models the second-order model is appropriate to describe the experimental data 
at the intermediate energy region over the energy of the resonance.
There has not been a way to include the width in the present formulation
because the divergences at $\omega = \alpha$ and $L$ in $Im \,T(\omega)$ are expressed by the function form 
different from the usual resonance formula of the Breit-Wigner type.
While the $\omega = \alpha$ one may simulate the (3,3)-resonance 
the latter one at $\omega = L$ moves $\sigma_\pm(\omega)$ to the opposite direction from each other.
It could be concerned with the resonance observed at the 0.7 $\sim$ 1 GeV region in $\sigma_{-}(\omega)$.
\\\hspace*{4.mm}
In Figure 1 and 2 the results of the total cross section $\sigma_\pm(\omega)$ by the second-order model 
are shown as a function of the laboratory momentum $p_{\rm L}$ for three values of the pseudovector coupling constant 
$f_\pi$ = 0.6, 0.7 and 0.8 of the interaction. 
It is a rather smaller than the standard value $f_\pi \sim 1$ extracted from the dispersion relation 
and the phase shift at the low-energy limit $\omega = m_\pi$ \cite{Haber-Schaim}.
As seen in the series of the theoretical curves the value of $f_\pi$ 
appropriate to the present method is around $f_\pi \sim 0.7$ although it changes by moving $L$.
The trend is similar to the calculation of the anomalous magnetic moment of nucleon. 
Inclusion of the self-energy in the scattering amplitude may be effective to enhance the value of $f_\pi$.
\\
\section*{\normalsize{4 \quad Summary }}
\hspace*{4.mm}
Taking the non-perturbative term into account
the pseudovector pion interaction enables us to calculate the higher-order corrections without the divergence.
The exact form of the self-energy does not necessarily provide the satisfactory results.
The approximate form is effective to explain the experiments of the pion-proton scattering at the intermediate energy region as well as the electromagnetic properties of nucleon.
It implies that the research of the method is in progress 
and applicable to phenomena in which the degree of the off-shell property is relatively weak.
The nucleon propagator for the intermediate state has not been corrected by the self-energy
which is expected to affect the scattering amplitude.  
\\
\hspace{4.mm}

\newpage
\begin{figure}
\begin{center}
\scalebox{0.5}{\includegraphics{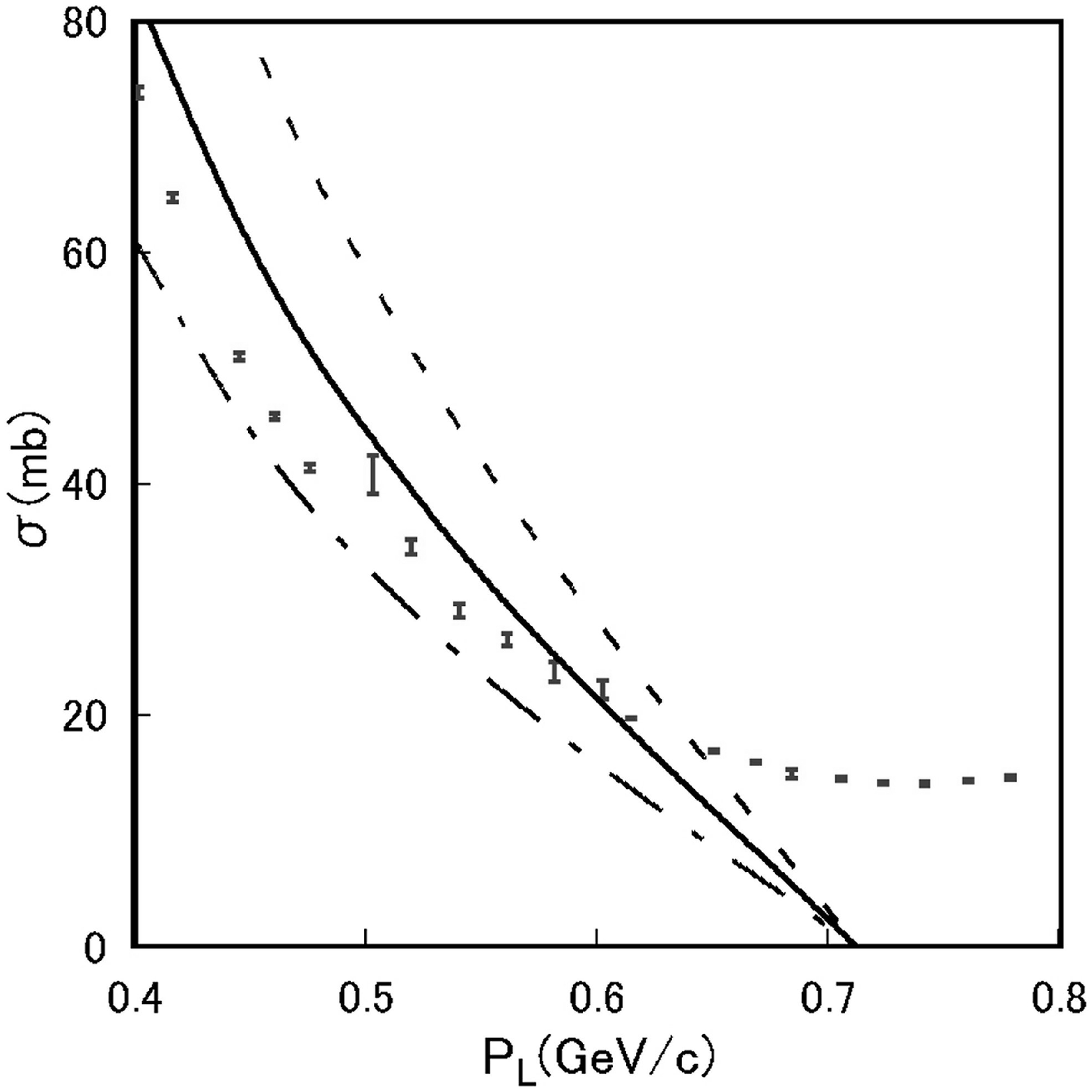}}
\caption{
The total cross section $\sigma$ (mb) of the $\pi^{+}$-proton scattering 
as a function of the laboratory momentum $p_L$ (GeV/c). 
Three curves are results of the calculation.
The dashed line ($f_\pi$=0.8).
The solid line ($f_\pi$=0.7). 
The dash-dot line ($f_\pi$=0.6).
The parameter of the periodicity is $L$=0.9.
The experimental data are taken from ref. \cite{PDG}.
}
\end{center}
\end{figure}
\newpage
\begin{figure}
\begin{center}
\scalebox{0.5}{\includegraphics{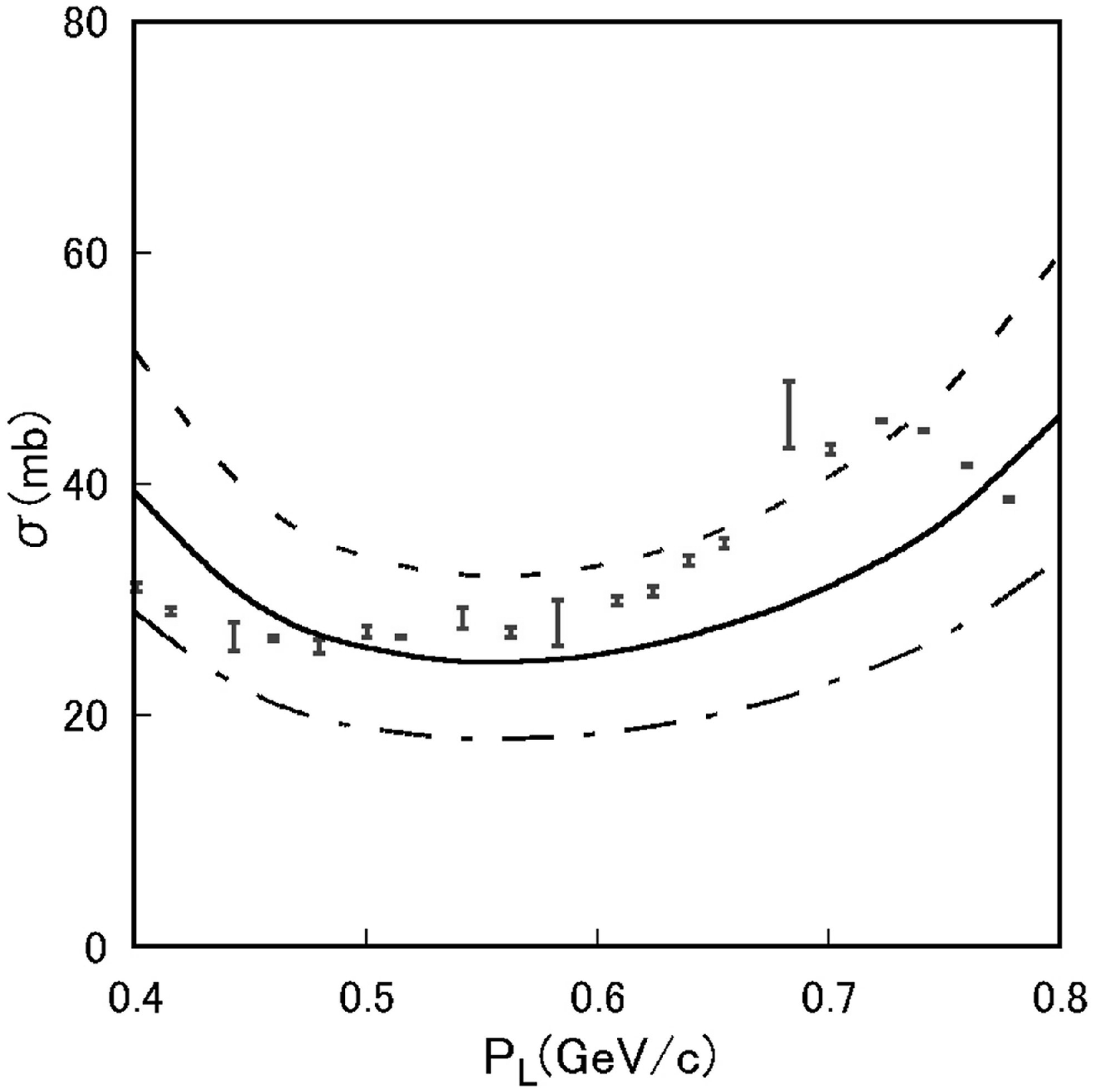}}
\caption{
The total cross section $\sigma$ (mb) of the $\pi^{-}$-proton scattering 
as a function of the laboratory momentum $p_L$ (GeV/c). 
Three curves are results of the calculation.
The dashed line ($f_\pi$=0.8).
The solid line ($f_\pi$=0.7). 
The dash-dot line ($f_\pi$=0.6).
The parameter of the periodicity is $L$=0.9.
The experimental data are taken from ref. \cite{PDG}.
}
\end{center}
\end{figure}
\end{document}